\documentclass[runningheads]{llncs}
\usepackage{graphicx}
\usepackage{booktabs}
\usepackage{graphicx}
\usepackage{subcaption}
\usepackage{multicol}

\begin{document}
%\title{Is a humorous robot more trustworthy? An experimental study in human-robot interaction
%
\title{Is a humorous robot more trustworthy?\thanks{This research was supported in part by a grant from the Priority Research Area DigiWorld PSP: U1U/P06/NO/02.19 under the Strategic Programme Excellence Initiative at the Jagiellonian University, and by the National Science Centre, Poland, under the OPUS call in the Weave programme under the project number K/NCN/000142.}}

\author{Barbara Sienkiewicz\inst{1}\orcidID{0009-0008-1977-5806} \and
Bipin Indurkhya\inst{1}\orcidID{0000-0002-3798-9209}}
\authorrunning{B. Sienkiewicz and B. Indurkhya}

\institute{Cognitive Science Department, Jagiellonian University, Krakow, Poland.
\email{barbara.wzietek@student.uj.edu.pl}\\
\email{bipin.indurkhya@uj.edu.pl}}
\maketitle              % typeset the header of the contribution
\begin{abstract}
    
As more and more social robots are being used for collaborative activities with humans, it is crucial to investigate mechanisms to facilitate trust in the human-robot interaction. One such mechanism is humour: it has been shown to increase creativity and productivity in human-human interaction, which has an indirect influence on trust. In this study, we investigate if humour can increase trust in human-robot interaction. We conducted a between-subjects experiment with 40 participants to see if the participants are more likely to accept the robot's suggestion in the Three-card Monte game, as a trust check task. Though we were unable to find a significant effect of humour, we discuss the effect of possible confounding variables, and also report some interesting qualitative observations from our study: for instance, the participants interacted effectively with the robot as a team member, regardless of the humour or no-humour condition. 

\end{abstract}


\section{Introduction}

In recent years, social robots are increasingly being deployed in various roles where they need to interact heavily with human users or play the role of a human. Some examples of such domains are military \cite{springer_military_2013}, medicine \cite{joshi_medical_2020}, consumer assistant \cite{ivanov_robot_2017}, healthcare, and care for the elderly \cite{schaefer2016meta}. However, to have an effective and meaningful interaction with the robot, it is necessary that people consider them trustworthy and reliable \cite{lee_review_2008}. 

This applies to all human-robot interactions, whether they take place on land, sea, air, or in cyberspace. Many factors can influence trust. In the research presented here, we explore the role of humour in facilitating trust in human-robot interaction (HRI, henceforth). Humour has been shown to increase creativity and productivity in human-human interaction; moreover it has an indirect influence on trust in humans. We conducted a study with 40 participants, using a between-subjects design with humor and no-humor conditions, to determine whether participants would be more inclined to accept a robot's suggestions in the Three-card Monte game, which we considered as a measure of trust towards the robot. In this paper,  we present the details of the pilot and the main studies, and discuss the results.

The structure of the paper is as follows. We present a review of the related research in Sec. 2. Then our experiment design and the pilot study is presented in Sec. 3, followed by the details of the main study in Sec. 4. Results and discussion are presented in Sec. 5, followed by the conclusions and suggestions for future research in Sec. 6.

\section{Related work}

We provide here background and motivation for this research. 
First, we introduce into the concept of trust in human-robot interaction (HRI), which is a key to effective collaboration between humans and robots, with an emphasis on the potential pitfalls of over-trust and under-trust. Next, we explore trust in human-human communication, shedding light on the dynamic nature of trust judgments and the factors influencing them. We then discuss the role of humor in human-human communication, emphasising its subjectivity and its impact on trust in human interactions. Then we briefly describe the existing research on the role of humor in HRI, focussing on its potential to enhance trust and likeability. This motivates our main research question: Does the use of humor in human-robot interaction improve trust?

\subsection{Trust in human-robot interaction}
Before we start discussing different aspects of trust, we need to define trust. Though there are many definitions of trust \cite{khavas_review_2021}, \cite{o2012general}, we use the following definition from \cite{nam_trust_2020}:

``Trust is a dyadic relation in which one person accepts vulnerability
because they expect that the other person's future action will have
certain characteristics; these characteristics include a mix of
performance (ability, reliability) and/or morality (honesty, integrity,
and benevolence).''

An inappropriate level of trust between a human user and a robot may lead to misuse or disuse of the robotic agent \cite{lee_trust_2004}. Misuse occurs when the user over-trusts the robot and accepts all its suggestions without questioning: for example trusting a GPS-based route assistant blindly while ignoring the actual situation on the road \cite{Hansen_2015}.
%An example of misuse is the blind trust in the GPS assistant, while ignoring actual situation on the road: the GPS assistant might not have the current information on the traffic.
On the other hand, disuse appears when the human user rejects all the suggestions of a robot and questions its capabilities: for example when a senior with Alzheimer’s disease does not believe the robot that she or he has not yet taken the medication. Both misuse and disuse undermine the effectiveness of human-robot interaction, and it is important to study how to maintain trust between a robotic system and its human user.

There are many factors that affect trust in HRI. Studies have shown that humans are more likely to trust robots that exhibit social cues, such as eye contact, appropriate facial expressions, and naturalistic movements \cite{mutlu2009footing}. Further, it has been found that people are more willing to trust a robot when it apologises and acknowledges that it made a mistake \cite{larzelere_dyadic_1980}. Another work of research has found that embodiment increases trust towards the robot \cite{moura_oliveira_progress_2019}. Trust in HRI can be divided into two categories: performance-based trust and relation-based trust \cite[p. 28-32]{nam_trust_2020}. Performance-based trust centres on the robot being capable and competent for its task: for example, in autonomous cars or banking systems. This trust is based on rational arguments and beliefs. Relation-based trust focuses on the robot's role as a social agent, which is more important when the robot serves as a companion, in a nursing home, or in a school. The users may not be completely aware why and how they trust the robot; they just feel more secure and comfortable about relying on the robotic system beyond the available evidence \cite{moussawi_effect_2020}. 

Many of the factors mentioned in the earlier research are rooted in interpersonal behaviours. Therefore, it is important to consider what influences trust in human-human interaction, as this may provide clues to facilitating trust in HRI.

\subsection{Trust in human-human communication}
Trust plays a key role in social interaction, and is crucial for effective cooperation. When two humans interact, whether they are strangers or close associates, each one has to decide how much to trust the other \cite{gladwell2020talking}. Furthermore, trust is dynamic: it changes as the interaction proceeds based on many conscious and subconscious factors. At first, we judge trustworthiness based on facial features \cite{willis2006first}. As we get to know each other better, this assessment of trustworthiness changes \cite{alarcon2016effect}. Sometimes, the social position of a person, like being a doctor, increases the initial assessment of trust in her or him. Another factor that leads humans to trust another is similarity to themselves. Similarities can include common values (such as strong work ethics), membership in defined groups (local churches and even gender), and common personality characteristics (extroversion and ambition) \cite{hurley_decision_2006}.

\subsection{Humour and trust in human-human communication}
Humour is an activity that is largely subjective and hard to measure objectively \cite{mcdonald2013philosophy}. 
Human brain is capable of taking into account many factors, such as the situation, atmosphere, and mood of the people around, to produce fun and bring smiles on people's faces. Everyone has a different sense of humour: some people like jokes, some gallows humour, and still others like pranks. Everyone reacts differently, even to the same joke. Due to the many variables and cultural conditioning, it is difficult to define humour precisely. For the this research, we consider a good sense of humour to be the ability to create jokes, riddles, and situations that make people laugh. Previous research has shown that humans with a good sense of humour are more intelligent \cite{greengross_humor_2011}. Most relevant to our research is the finding that humour can increase trust between agents \cite{kim_supervisor_2016}. In teams, humour is found to increase productivity and creativity, reduce conflict, and decrease stress \cite{holmes_making_2007}.

\subsection{Research on the role of humour in HRI}
Given that humour has such a positive influence on human-human interaction, it is useful to explore if it can also facilitate human-robot interaction. Anton Nijholt studied the specific role and use of humour in human-computer interaction, and demonstrated the potential of using humour with special attention on humour creation\cite{nijholt2017humor}. Another study found that humour helps robots to interact with humans in the same way as humans interact with each other \cite[p. 333-360]{hempelmann2008cartoons}. Yet another study showed that the interaction with humour is more natural and flexible \cite{mulder2002humour}, and it might be correlated with trust. Humour can also help robots to be perceived as smarter: for example, when they tell clever jokes \cite{inproceedings}. Francesco Vigni et al. argue that users perform better in a game when interacting with an agreeable robot \cite{vigni2023sweet}."

An empirical study of humour with Nao and iCat robots demonstrating different laughing behaviours showed higher likeability ratings for the robots when they use humour \cite{mirnig2016robot}; interestingly, the likeability ratings tended to converge when either robot laughed or when both robots laughed together. This suggests that it is not necessary for both participants in the interaction to have a sense of humour in order to yield a positive effect of humour, which is important as we are not able to manipulate the human's humour in the interaction. 
Another study on humour in HRI showed that using jokes during the initial greeting is effective in enhancing likeability and reducing awkwardness \cite{tae2020effect}. Our methodology in creating the study presented here is inspired by it.  

Based on the existing literature, there is evidence to suggest that humour can indirectly affect trust by increasing the likeability and naturalness of interactions between humans and robots. However, the question remains whether the use of humour in HRI can also directly enhance trust between humans and robots. Therefore, the research question we explore in this study is: Does the use of humour in human-robot interaction improve trust?

\section{Experiment Design}

In this section, we present the main task for the participants in the experiment, introduce the robot and its software, and provide a detailed description of the obstacles we encountered during the pilot study.

\subsection{Task}

To investigate whether people consider a humorous robot more trustworthy, we designed an experiment where the participant is asked to team up with a robot while playing a version of the card game {\it Three-card Monte.} In our version of the game (Fig. 1b), the participant is shown three cards, two of which are black and one red. The cards are then turned face down and shuffled out of the sight of the participant and the robot (behind the back of the experimenter). 
The experiment was deliberately designed so that the robot did not have more information than the participant. This allows us to focus on relation-based trust, when both agents have the same information, and there is no rationale for trusting the robot.

The three cards are shown to the participant again, who is asked to point to the red card. The robot then offers a suggestion that is contrary to the participant’s guess. We observe if the participant follows its advice, thereby showing that she or he trusts the robot: this measure is based on the experimental types for measuring trust in the book 'Trust in Human-Robot Interaction' \cite{nam_trust_2020}. 

We also use the Multi-Dimensional Measure of Trust (MDMT) questionnaire after the interaction to measure the subjective level of trust towards the robot \cite{malle_multi-dimensional_2019}.

\begin{figure}[h]
    \begin{subfigure}{0.35\textwidth}
        \centering
        \includegraphics[width=\linewidth]{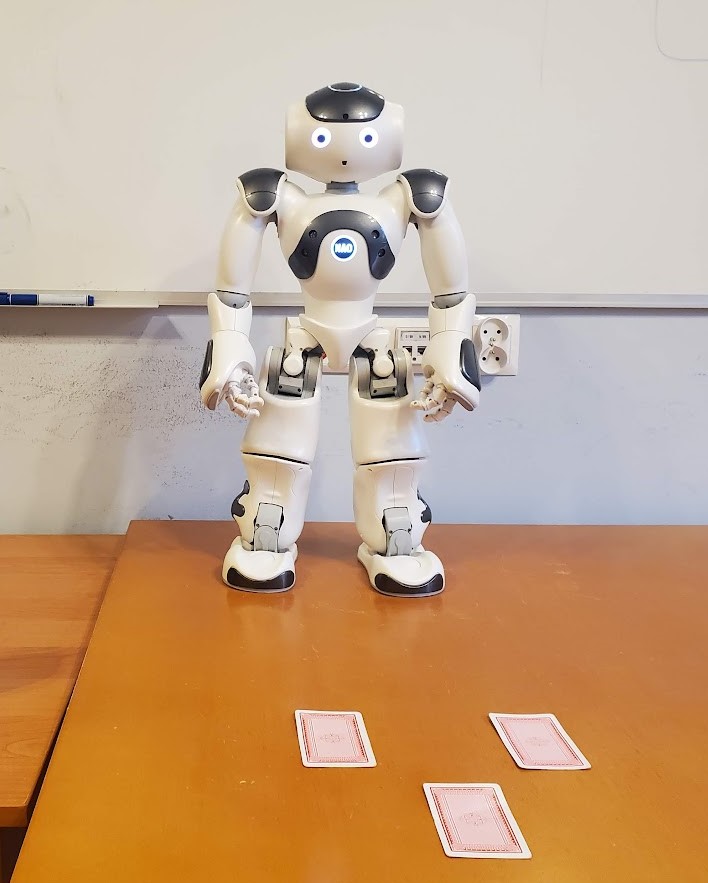}
        \caption{Nao and the cards }
        \label{Fig:NaoRobotCardGame}
    \end{subfigure}
    \begin{subfigure}{0.60\textwidth}
        \centering
        \includegraphics[width=\linewidth]{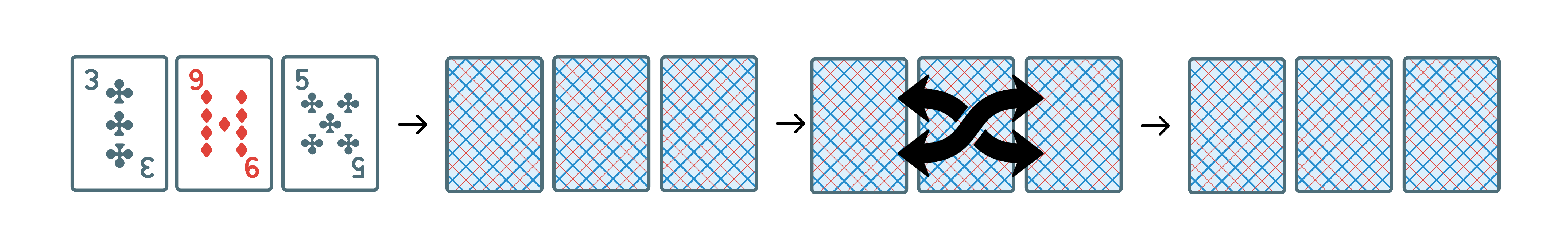}
        \caption{Game order}
        \label{Fig:ThreeCardMonte}
    \end{subfigure}
    \vfill
    \caption{Main task: Three-card Monte}
\end{figure}
To study the effect of humour, we compared humour and no-humour conditions. Three jokes were selected to be used in the humour (experimental) condition. All three jokes employed a form of wordplay in the Polish language. For example, one of the jokes used during the interaction is as follows: "What is the name of the cat that flies?" The punchline to this joke relies on a wordplay between the Polish words for cat ("KOT") and flies ("LECI"), resulting in the humorous response of "small cutlet" ("KOTLECIK").

It is important to note that the jokes were chosen to be neutral with respect to factors such as race, gender, and profession, in order to minimize the potential for offense or biases. The same jokes were used for the pilot study as for the main study. Additionally, one situational joke was included in the main study, where the robot humorously remarks, "Sweets are just for you, I'm on a diet."

In the no-humour (control) condition of the experiment, the robot engaged in a conversation with the participant about some neutral topic like the weather, before proceeding to the main task. This serves as a baseline for comparison against the conditions involving joke interactions.

Throughout the experiment, the same pre-selected jokes were consistently used during the interactions, with an additional joke available in case the participant expressed interest in hearing more.

\subsection{Robot}

We used the humanoid robot NAO from Aldebaran (Fig. 1a). Nao is 58 cm tall and has 25 degrees of freedom. The same NAO robot was used in both conditions. We used the methodology of Wizard-of-Oz \cite{riek2012wizard}. The robot’s voice and talking speed were also kept identical. Nao turned its face towards sounds, which made the interaction more natural when we modified the experimental setting after the pilot study, as explained below. 

\subsection{Pilot study}
We first conducted a pilot study with 12 participants to test our methodology. The study was conducted in the Social Robotics Lab at Jagiellonian University in Krakow, Poland. In this pilot study, we included another condition, namely using a video of the robot instead of an embodied robot. This led to a between-participants design where each participant was tested with one of the following four conditions: 1) A humorous robot; 2) a neutral robot; 3) a humorous robot displayed on a tablet; and 4) a neutral robot displayed on a tablet. We list below some problems observed in the pilot study and how they were fixed for the main study. 

\begin{description}
\item [Rigid and unnatural movements of the robot:]
The robot moved unnaturally and stiffly, which resulted in some participants failing to notice it at all and talking only to the experimenter. To remedy this problem, we implemented animated speech for the main study, so the robot was constantly moving, gesturing, and making small talk (initiated by the operator), which did not depend on the input from the robot’s operator. 
\item [Slow and delayed speech from the robot:]
To address slow speech and typing sounds, we pre-programmed certain robot responses, assigning each an index. During interactions, the operator only needed to input the corresponding index, allowing for quicker and more seamless communication. The operator could still type responses in real time for unexpected questions or comments from participants.
\item [Unnatural communication with the avatar:]
Two conditions required the participants to talk to the avatar on the tablet and read messages on the screen. They found this interaction quite weird and did not feel comfortable. So, we decided to remove this condition (interaction with an avatar) from the main study.  
\item [Lack of eye contact:] 
In the pilot study, the robot remained stationary but faced away from the participant while speaking due to the operator's limited view. This felt unnatural to participants. In the main study, we improved this by giving the operator a view of both the participant and the robot, using a second computer to control the robot's head orientation for better interaction.
\end{description}

\section{Main study}\label{AA}
The main experiment was also carried out in the Social Robotics Lab at Jagiellonian University in Krakow, Poland. It was organized as a between-participants study, with each participant being assigned to one of the two conditions: 1) a humorous robot (experimental condition); and 2) a neutral robot (control condition).

\subsection{Participants}
Forty participants (F=17, M=22, Non-binary =1; Mean age = 28,97; SD=10,71) were recruited through social media and departmental email. During the recruitment, participants were asked if they ever met a humanoid robot (two participants had taken part in a previous study, so had met a humanoid robot before) and about their background (age, sex, educational background). The participants were randomly assigned to the experimental group or the control group. Each participant was paid 20 PLN (about 5 Euros) in cash for their participation. 

\subsection{Experimental set up}
The setup is shown in Fig. 3. The light in one part of the laboratory was turned off to hide the robot operator, and most participants did not see the operator. The participants were told at the beginning of the experiment that the robot is autonomous, and they did not show any indication of doubting this assumption. 

The robot was operated by the same person for all the participants. The operator could see the participant through the camera on the robot's forehead (Fig. 4). The robot was stationary throughout the experiment but moved its head and arms naturally (live mode). All the conversation with the participant was conducted by the operator through the robot. The operator used two computers to control the robot: one running the Choregraphe program to control the robot’s body movements and the view of the participant, and the other to control the robot’s speech. 

While filling out the questionnaire, the participant and the experimenter were near the door (position 1), during the initial small-talk phase, when the robot told jokes or commented on the weather, the participant was in position 2 and the experimenter in position 1, and during the game, both the participant and the experimenter were in positions 2. The interaction with the robot was recorded with the participant’s consent.  

\begin{figure}[h]
    \begin{subfigure}{0.4\textwidth}
        \centering
        \includegraphics[width=\textwidth]{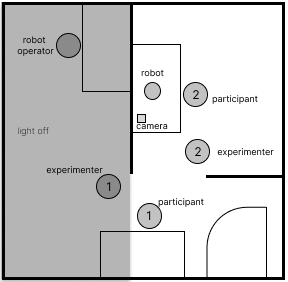}
        \caption{Schematic of the study setup}
        \label{Fig:SchematicStudySetup}
    \end{subfigure}
    \hfill
    \begin{subfigure}{0.4\textwidth}
        \centering
        \includegraphics[width=\textwidth]{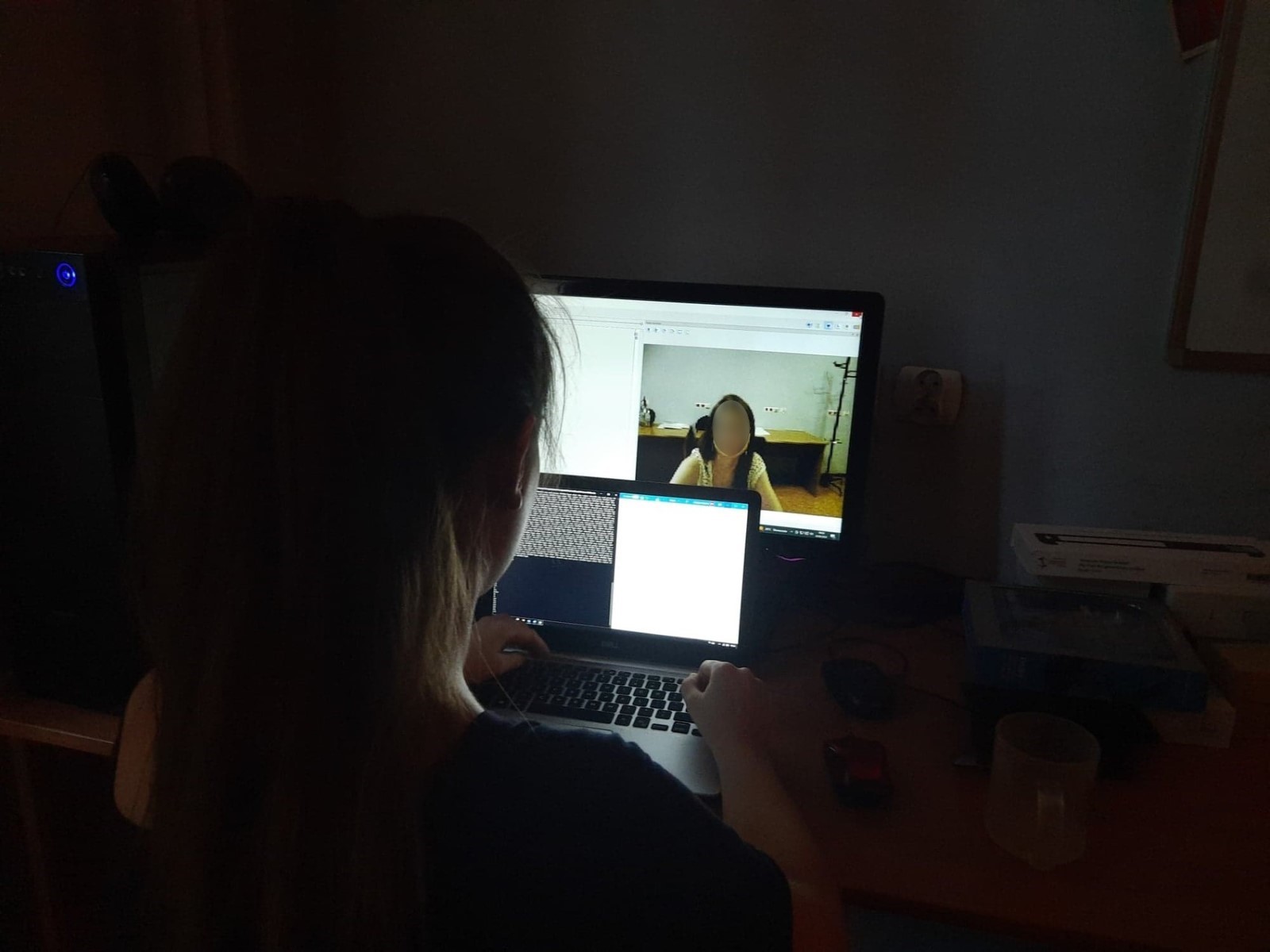}
        \caption{View of the robot’s operator during the experiment.}
        \label{Fig:RobotOperatorView}
    \end{subfigure}
    \caption{Experiment set up}
    \label{Fig:TwoImagesSideBySide}
\end{figure}
\subsection{ Procedure }
The participant entered the lab and was welcomed by the experimenter. Then they sat down at the table,
and the participant was asked to sign the informed consent form. The participant was given a paper with the introduction to the experiment and the instructions. Then the participant was led to sit in front of NAO and wait for it to start the conversation. NAO welcomed the participant, introduced itself, and asked the participant’s name and how he or she is. Then, for the experimental group, NAO made some small talk and told two jokes to the participant. The conversation always had the same structure, the jokes came at the same time for each participant and were only in this part of the experiment. For the control group, NAO made some remarks about the weather.

After the conversation, NAO explained the rules of the game, and then the researcher entered the space, showed the cards, turned them face down, and shuffled them out of the participant’s sight. The researcher then spread the cards on the table face down, and asked the participant, ‘Where is the red card?’ After the participant made a guess, NAO suggested a card different than what the participant had chosen. If the participant asked why, NAO replied that it was its intuition. The researcher then stopped further interaction and asked the participant to choose a card without flipping it. The participant was asked to return to the first table and fill out the questionnaire. The researcher then let the participant reveal the card. Regardless of the colour of the card, the participant was offered a sweet as a prize. The participant was then given the participation fee. Finally, the robot operator was exposed and, depending on the interest of the participants, the study was explained, and the experimenter answered any questions. The entire procedure took 15 to 20 minutes.  

\subsection{ Questionnaire }
We used the Multi-Dimensional Measure of Trust (MDMT) questionnaire to measure the subjective level of trust in the robot [19]. The MDMT contains 16 elements to assess four differentiable trust dimensions. An agent can be trusted because it is reliable, competent, ethical, and/or sincere. These four dimensions are organized into two broader trust factors: Capacity Trust (Reliable, Capable) and Moral Trust (Ethical, Sincere). Moral trust is the subjective level of trust a person has in the robot's ethical and reliable behavior, while capacity trust is the subjective level of trust a person has in the robot's abilities or performance in executing its designated tasks. 

Each of the 16 items was to be evaluated on an 8-point Likert scale from 0 (not at all) to 7 (very). They could also choose the option ‘Does not fit’, to avoid a forced answer. This questionnaire is widely used to measure trust in HRI and for comparison with other research. The original version is in English, which we translated into Polish for our study.  

\section{Results and Discussion}
\subsection{Objective independent variable}
As shown in Table 1, about 75\% of the participants followed the robot’s advice. We conducted a Chi-square test to assess whether the proportion of participants who changed their decision differed significantly between the humour and the non-humour conditions. We did not find a significant difference (p = 0.288).

\begin{table}[h]
  \caption{Proportion of Participants Who Trusted the Robot}
  \label{tab1}
  \centering
  \begin{tabular}{lcc}
    \toprule
    \textbf{Condition} & \textbf{\textit{Followed the robot}} & \textbf{\textit{Did not follow the robot}} \\
    \midrule
    Humourous Nao & 13 & 7 \\
    Non-humorous Nao & 16 & 4 \\
    \midrule
    \textbf{\textit{Total}} & \textbf{29} & \textbf{11} \\
    \bottomrule
  \end{tabular}
\end{table}
 \begin{figure}[h]
\centerline{\includegraphics[width = 0.4\textwidth] {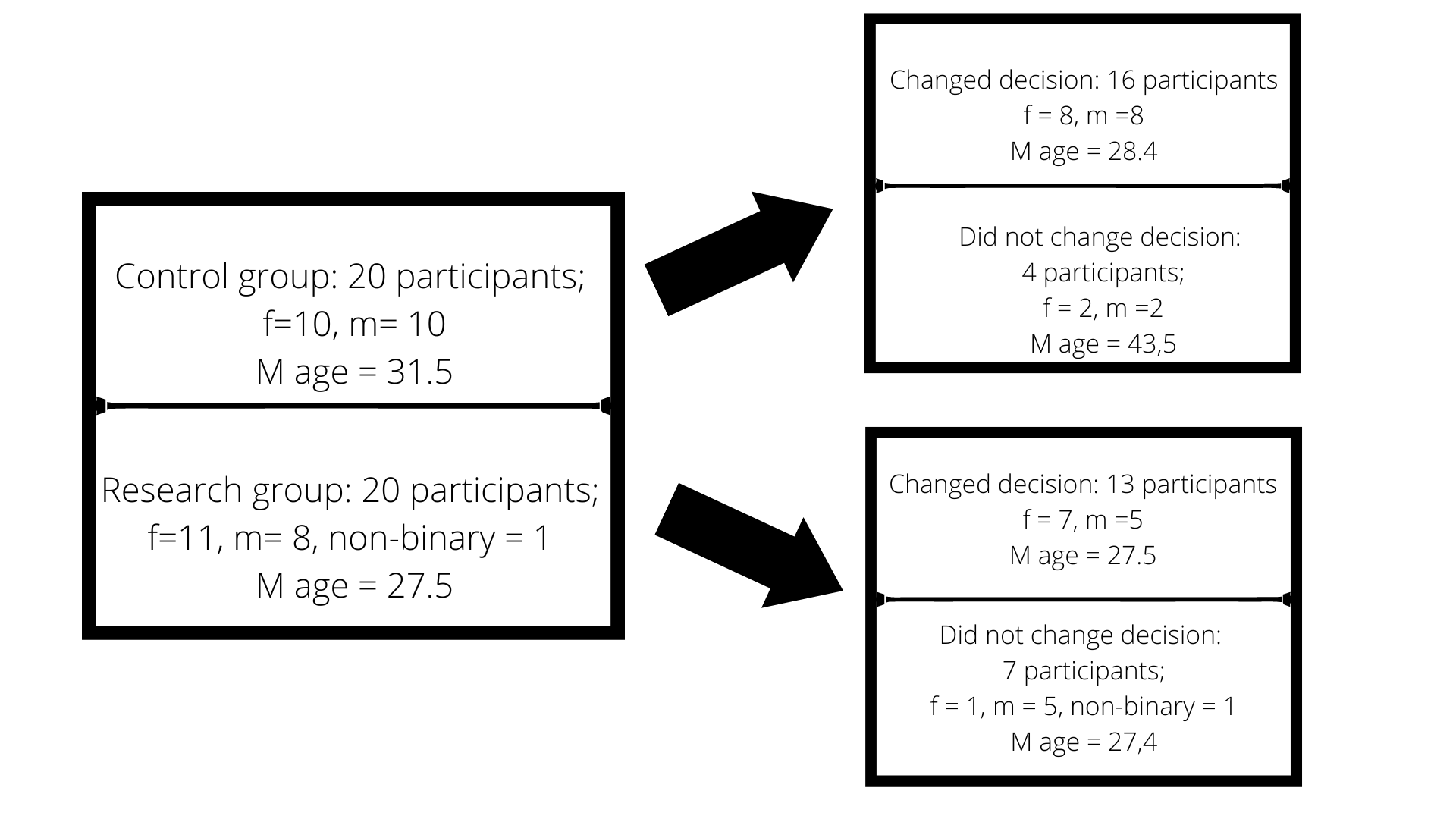}}
\caption{Effect of gender and age in humour and non-humour conditions}
\label{Fig. Gender and age distribution in each group }
\end{figure} 
For representativeness of the sample, we collected the age and gender of the participants before the study. We did additional analysis for the effect of age and gender on whether the participants accepted the suggestion by the robot in both the humour and the non-humour conditions (Fig. 5). We expected that more participants in the non-humour group would not accept the robot's suggestion, but we found that fewer participants (4 out of 20) did not change their decision. However, the mean age of these participants (43.5 yrs) was significantly higher than the other group (28.4 yrs). Though there were too few participants to make any generalizations (no statistical significance), this observation is consistent with the previous research that shows that older adults have a more negative attitude towards robots than younger adults \cite{chien_age_2019}. We also noted that in the humour group, seven out of twenty participants did not accept the robot's suggestion, and only one of them described herself as a female. The present observation is in agreement with the prior study, which has suggested that women tend to place more trust in robots as compared to men \cite{gallimore2019trusting}.

 Moreover, it is worth noting that two participants from the non-humorous group changed their  decision to select a card that was neither suggested by the robot nor their initial preference. They might have chosen to deviate from the robot's recommendation while also wanting to avoid any potential conflict within the group. This possibility suggests that social dynamics and the desire to maintain positive relations within the team may have played a role in their decision-making process.

It should be emphasized that our study's outcomes may have been influenced by certain biases that could be attributed to both the nature of the task and the participants' prior knowledge of robots. For example, the experimenter shuffled the cards behind her back, which meant that the participant did not know where the red card was and had to guess. This might have created a negative attitude toward the task. In contrast, in the standard version of Three-card Monte, the three cards are moved around rapidly in the plain view of the participant. 
Moreover, the participants did not have any information about how the robot worked and what basis, if any, it might have to make its suggestion. Some participants reported after the study that theirs was a pure guess, that they decided to trust the robot because they thought it might have more information.

Another variable affecting the results is the participants' varying sense of humour. We prepared three jokes, two for the interaction and one just in case. The jokes were neutral according to race, gender, profession, and other factors that might be considered offensive; there was also one situational joke. However, participants could have had a different sense of humour and may not have liked the robot’s humour.  

\subsection{Subjective independent variable}
We conducted the Mann-Whitney U test on the results of the MDMT questionnaire to see if there was any difference between the humour and the non-humour conditions (Table II). However, we did not find a significant difference either for the capacity trust (p=0.465) or for the moral trust (p=0.378).

\begin{table}[t]
  \caption{Average scores in MDMT}
  \label{tab:mdmt}
  \centering
  \begin{tabular}{lccc}
    \toprule
    \textbf{Condition} & \textbf{\textit{Capacity trust}} & \textbf{\textit{Moral trust}} & \textbf{\textit{Trust}} \\
    \midrule
    Control condition & 4.76 & 5.49 & 5.12 \\
    Research condition & 4.98 & 5.70 & 5.34 \\
    \bottomrule
  \end{tabular}
\end{table}

Although we did not get significant differences in the Mann-Whitney U test, average scores in MDMT questionnaire had recurring differences. The participant who interacted with the humorous robot gave higher responses in the MDMT questionnaire than in the control group. This is visible on the capacity sub-scale and also in the moral sub-scale. The average score on the moral sub-scale trust is higher than the capacity trust in both conditions, suggesting relation-based trust. We also found a strong correlation between those sub-scales (r(39) $= .794$; $p < .01$), which explains the recurring differences. 

It is worth examining the genesis of the questionnaire we used. The MDMT questionnaire was constructed by researchers who conducted open surveys and questionnaires with people recruited through Amazon Mechanical Turk. In one part of the creation, sorting trust words, participants were asked to consider 32 words or short phrases, 6–7 for each of the four hypothesized dimensions, as well as five filler items assumed to be unrelated to trust, which included the word `humourous’ [7]. Thus, this questionnaire assumed that humour is not related to trust. The MDMT questionnaire was constructed not for interaction with robots, but to distinguish different dimensions in trust, according to human, institutions, or robots. However, as far as we are aware, no studies have been conducted on humour and trust between humans with the MDMT questionnaire, so this area is still unexplored.  

\subsection{Relation of the objective and subjective variables }
We conducted a Mann-Whitney U test to see if the MDMT questionnaire results correspond to the decision to accept or not accept the robot's suggestion during the card game. We expected that the participants who changed their decision rated the robot higher in the MDMT questionnaire. However, we did not get a significant result in this regard ($p > 0.05$). This may be due to different personal attitudes of the participants, which requires further study \cite [p.25]{nam_trust_2020}. Some participants trust robot less even if they accept the robot's suggestion, because their general baseline trust is lower.    

\subsection{Qualitative analysis: behavioural observations}
Analyzing the recording from the main study, we noticed some interesting behaviour patterns of the participants. For example, at the beginning of the interaction with the robot, many participants had their hands hidden under the table. But as time passed, and before they were asked to choose a card, they put their hands on the table. This suggests that these participants felt more comfortable with the robot as the interaction proceeded, and did not feel the need to keep a safe distance. (We excluded the possibility that the participants were cold as it was summer in Poland then, and the experimental room was at a comfortable temperature.)

Another observation was that some participants asked the robot why it thinks that the other card is red. Even though these participants asked for an explanation but did not receive it (the robot answered `That is my intuition'), most of them followed the robot's suggestion.

Finally, a surprising behaviour was observed from most participants at the end of the interaction. After the participant selected the card, the experimenter asked them to go back to the table and complete the final questionnaire. As the participant stood up to leave, the robot conveyed its final message, ``Thank you for the game, and see you later!'' The participants consistently responded with a farewell remark such as ``Thank you too, bye'' or ``It was nice to meet you''. This suggests that the participants treated the robot as a social agent, a team member, or even a partner. This observation adds to the current understanding of human-robot interaction and highlights the need for further exploration of this phenomenon.
%Therefore, this result serves as a valuable contribution to the broader understanding of the dynamics involved in human-robot interactions.

\section{Conclusion}

The primary objective of this study was to investigate the influence of humor on trust in the context of human-robot interaction (HRI). Despite not being able to confirm the main hypothesis, our study found some valuable insights by implementing a between-subjects study design. Our findings demonstrate that participants consistently engaged in effective teamwork with the robot, regardless of the presence or absence of humor.

Our study provides a notable contribution to the field of Human-Robot Interaction (HRI) by introducing a framework for using the Wizard-of-Oz methodology. This approach facilitated a controlled manipulation of humor in the robot's behavior while ensuring a realistic and interactive environment for participants. The pilot study played a crucial role in refining our methodology and improving the overall quality of the study. We believe that the comprehensive documentation of challenges encountered during the study serves as a valuable contribution to the field by highlighting potential pitfalls for future research endeavors.

The effect of humor on trust in human-robot interaction needs more research overcoming our limitations to 
be able to make further conclusions.

\section*{Acknowledgment}
\vspace{-2mm}
We thank Anna Kołbasa and Sharon Spisak for their help in conducting this study.

\bibliographystyle{splncs04}
\bibliography{main.bib}

\begin{thebibliography}{10}
\providecommand{\url}[1]{\texttt{#1}}
\providecommand{\urlprefix}{URL }
\providecommand{\doi}[1]{https://doi.org/#1}

\bibitem{alarcon2016effect}
Alarcon, G.M., Lyons, J.B., Christensen, J.C.: The effect of propensity to trust and familiarity on perceptions of trustworthiness over time. Personality and Individual Differences  \textbf{94},  309--315 (2016)

\bibitem{chien_age_2019}
Chien, S.E., Chu, L., Lee, H.H., Yang, C.C., Lin, F.H., Yang, P.L., Wang, T.M., Yeh, S.L.: Age {Difference} in {Perceived} {Ease} of {Use}, {Curiosity}, and {Implicit} {Negative} {Attitude} toward {Robots}. ACM Transactions on Human-Robot Interaction  \textbf{8}(2),  9:1--9:19 (Jun 2019). \doi{10.1145/3311788}, \url{https://doi.org/10.1145/3311788}

\bibitem{gallimore2019trusting}
Gallimore, D., Lyons, J.B., Vo, T., Mahoney, S., Wynne, K.T.: Trusting robocop: Gender-based effects on trust of an autonomous robot. Frontiers in Psychology  \textbf{10}, ~482 (2019)

\bibitem{gladwell2020talking}
Gladwell, M.: Talking to strangers. Gramedia Pustaka Utama (2020)

\bibitem{greengross_humor_2011}
Greengross, G., Miller, G.: Humor ability reveals intelligence, predicts mating success, and is higher in males. Intelligence  \textbf{39}(4),  188--192 (Jul 2011). \doi{10.1016/j.intell.2011.03.006}, \url{http://www.sciencedirect.com/science/article/pii//S0160289611000523}, publisher: JAI

\bibitem{Hansen_2015}
Hansen, L.: 8 drivers who blindly followed their gps into disaster (Jan 2015), \url{https://theweek.com/articles/464674/8-drivers-who-blindly-followed-gps-into-disaster}

\bibitem{hempelmann2008cartoons}
Hempelmann, C.F., Samson, A.C.: Cartoons: drawn jokes? The primer of humor research pp. 609--640 (2008)

\bibitem{holmes_making_2007}
Holmes, J.: Making {Humour} {Work}: {Creativity} on the {Job}. Applied Linguistics  \textbf{28}(4),  518--537 (Dec 2007). \doi{10.1093/applin/amm048}, \url{https://academic.oup.com/applij/article-lookup/doi/10.1093/applin/amm048}

\bibitem{hurley_decision_2006}
Hurley, R.F.: The decision to trust. Harvard business review  \textbf{84}(9),  55--62 (2006)

\bibitem{ivanov_robot_2017}
Ivanov, S.H., Webster, C.: The {Robot} as a {Consumer}: {A} {Research} {Agenda}. {SSRN} {Scholarly} {Paper} ID 2960824, Social Science Research Network, Rochester, NY (2017), \url{https://papers.ssrn.com/abstract=2960824}

\bibitem{joshi_medical_2020}
Joshi, S., de~Visser, E.J., Abramoff, B., Ayaz, H.: Medical {Interviewing} with a {Robot} {Instead} of a {Doctor}: {Who} do {We} {Trust} {More} with {Sensitive} {Information}? In: Companion of the 2020 {ACM}/{IEEE} {International} {Conference} on {Human}-{Robot} {Interaction}. pp. 570--572. ACM, Cambridge United Kingdom (Mar 2020). \doi{10.1145/3371382.3377441}, \url{https://dl.acm.org/doi/10.1145/3371382.3377441}

\bibitem{khavas_review_2021}
Khavas, Z.R.: A {Review} on {Trust} in {Human}-{Robot} {Interaction} (May 2021), \url{http://arxiv.org/abs/2105.10045}, arXiv:2105.10045 [cs]

\bibitem{kim_supervisor_2016}
Kim, T.Y., Lee, D.R., Wong, N.Y.S.: Supervisor {Humor} and {Employee} {Outcomes}: {The} {Role} of {Social} {Distance} and {Affective} {Trust} in {Supervisor}. Journal of Business and Psychology  \textbf{31}(1),  125--139 (Mar 2016). \doi{10.1007/s10869-015-9406-9}, \url{http://link.springer.com/10.1007/s10869-015-9406-9}

\bibitem{larzelere_dyadic_1980}
Larzelere, R.E., Huston, T.L.: The {Dyadic} {Trust} {Scale}: {Toward} {Understanding} {Interpersonal} {Trust} in {Close} {Relationships}. Journal of Marriage and the Family  \textbf{42}(3),  595--604 (Aug 1980)

\bibitem{lee_review_2008}
Lee, J.D.: Review of a {Pivotal} {Human} {Factors} {Article}: "{Humans} and {Automation}: {Use}, {Misuse}, {Disuse}, {Abuse}". Human Factors: The Journal of the Human Factors and Ergonomics Society  \textbf{50}(3),  404--410 (2008)

\bibitem{lee_trust_2004}
Lee, J.D., See, K.A.: Trust in {Automation}: {Designing} for {Appropriate} {Reliance}. Human Factors p.~31 (2004)

\bibitem{malle_multi-dimensional_2019}
Malle, B.F., Ullman, D.: A {Multi}-{Dimensional} {Conception} and {Measure} of {Human}-{Robot} {Trust} p.~21 (2019)

\bibitem{mcdonald2013philosophy}
McDonald, P.: The philosophy of humour. Humanities-Ebooks (2013)

\bibitem{mirnig2016robot}
Mirnig, N., Stadler, S., Stollnberger, G., Giuliani, M., Tscheligi, M.: Robot humor: How self-irony and schadenfreude influence people's rating of robot likability. In: 2016 25th IEEE International Symposium on Robot and Human Interactive Communication (RO-MAN). pp. 166--171. IEEE (2016)

\bibitem{moura_oliveira_progress_2019}
Moura~Oliveira, P., Novais, P., Reis, L.P. (eds.): Progress in {Artificial} {Intelligence}: 19th {EPIA} {Conference} on {Artificial} {Intelligence}, {EPIA} 2019, {Vila} {Real}, {Portugal}, {September} 3–6, 2019, {Proceedings}, {Part} {II}, Lecture {Notes} in {Computer} {Science}, vol. 11805. Springer International Publishing, Cham (2019). \doi{10.1007/978-3-030-30244-3}, \url{http://link.springer.com/10.1007/978-3-030-30244-3}

\bibitem{moussawi_effect_2020}
Moussawi, S., Benbunan-Fich, R.: The effect of voice and humour on users’ perceptions of personal intelligent agents. Behaviour \& Information Technology pp. 1--24 (Jun 2020). \doi{10.1080/0144929X.2020.1772368}, \url{https://www.tandfonline.com/doi/full/10.1080/0144929X.2020.1772368}

\bibitem{mulder2002humour}
Mulder, M.P., Nijholt, A.: Humour research: State of the art. Centre for Telematics and Information Technology, University of Twente (2002)

\bibitem{mutlu2009footing}
Mutlu, B., Shiwa, T., Kanda, T., Ishiguro, H., Hagita, N.: Footing in human-robot conversations: how robots might shape participant roles using gaze cues. In: Proceedings of the 4th ACM/IEEE international conference on Human robot interaction. pp. 61--68 (2009)

\bibitem{nam_trust_2020}
Nam, C.S., Lyons, J.B.: Trust in {Human}-{Robot} {Interaction}. Academic Press (Nov 2020), google-Books-ID: R8DvDwAAQBAJ

\bibitem{nijholt2017humor}
Nijholt, A., Niculescu, A.I., Valitutti, A., Banchs, R.E.: Humor in human-computer interaction: a short survey. Adjunct Proceedings of INTERACT pp. 527--530 (2017)

\bibitem{o2012general}
O'Hara, K.: A general definition of trust. \url{https://eprints.soton.ac.uk/341800/} (2012)

\bibitem{riek2012wizard}
Riek, L.D.: Wizard of oz studies in hri: a systematic review and new reporting guidelines. Journal of Human-Robot Interaction  \textbf{1}(1),  119--136 (2012)

\bibitem{schaefer2016meta}
Schaefer, K.E., Chen, J.Y., Szalma, J.L., Hancock, P.A.: A meta-analysis of factors influencing the development of trust in automation: Implications for understanding autonomy in future systems. Human factors  \textbf{58}(3),  377--400 (2016)

\bibitem{springer_military_2013}
Springer, P.J.: Military {Robots} and {Drones}: {A} {Reference} {Handbook}. ABC-CLIO (2013), {ISBN:} 978-1-59884-732-1

\bibitem{tae2020effect}
Tae, M., Lee, J.: The effect of robot's ice-breaking humor on likeability and future contact intentions. In: Companion of the 2020 ACM/IEEE International Conference on Human-Robot Interaction. pp. 462--464 (2020)

\bibitem{vigni2023sweet}
Vigni, F., Andriella, A., Rossi, S.: Sweet robot o’mine - how a cheerful robot boosts users' performance in a game scenario. RO-MAN Conference  (2023)

\bibitem{willis2006first}
Willis, J., Todorov, A.: First impressions: Making up your mind after a 100-ms exposure to a face. Psychological science  \textbf{17}(7),  592--598 (2006)

\bibitem{inproceedings}
Zhang, H., Yu, C., Tapus, A.: Why do you think this joke told by robot is funny? the humor style matters. pp. 572--577 (08 2022). \doi{10.1109/RO-MAN53752.2022.9900515}

\end{thebibliography}
\end{document}